\long\def\UN#1{$\underline{{\vphantom{\hbox{#1}}}\smash{\hbox{#1}}}$}
\def\NP{\vfil\eject}
\def\NI{\noindent}

\magnification=\magstep 2
\overfullrule=0pt
\hfuzz=16pt
\voffset=0.0 true in
\vsize=8.8 true in
\baselineskip 20pt
\parskip 6pt
\hoffset=0.1 true in
\hsize=6.3 true in
\nopagenumbers
\pageno=1
\footline={\hfil -- {\folio} -- \hfil}
 
\hphantom{A} 

\centerline{\UN{\bf Quantum signal splitting that avoids initialization
of the targets}}
 
\vskip 0.4in
 
\centerline{{\bf Dima Mozyrsky}\ and\ {\bf Vladimir Privman}}
 
\centerline{\sl Department of Physics, Clarkson University,
Potsdam, NY 13699--5820}
 
\vfill
 
\centerline{\bf ABSTRACT}
 
The classical signal splitting and copying are not possible
in quantum mechanics. Specifically, one cannot copy the 
basis up and down states
of the input ($I$) two-state system (qubit, spin) into
the copy ($C$) and duplicate-copy ($D$) two-state systems if the
latter systems are initially in an arbitrary state. We
consider instead a quantum
evolution in which the basis states of $I$ at time $t$ are
duplicated in {\it at
least two\/} of the systems
$I$, $C$, $D$, at time $t+\Delta t$. In essence, the restriction on 
the initial target states is exchanged for uncertainty as to which 
two of the three qubits retain copies of the initial source state.

\vfill
 
\NI {\bf PACS numbers:}\ 03.65.Bz, 85.30.St
 
\NP

\NI {\bf 1. Introduction and Definition of the Model}

\ 

The ``classical'' signal-copying process starts from the input
value $I$ and after some time
$\Delta t$ results in the same value at the
copy $C$ and, if needed, duplicate-copy $D$.
We assume that the value of $I$ is unchanged.
This is the case when a signal is copied, for instance,
by connecting wires and forcing the voltage in one of them to the
value 0 or 1. This input-wire voltage, and the equilibrium state,
will be established in all
the connected wires, after a time $\Delta t$ determined by the
relaxation processes of the charge-carrier
distribution in the wires. The important point to note is that 
this ``classical'' copying/duplicating of a signal is not governed by
reversible dynamics; there are inevitably some irreversible
dissipation processes involved.

Quantum-mechanical copying from $I$ to $C$, and more complicated,
multi-copy processes, have been discussed in the literature [1-6].
Generally, one cannot copy an arbitrary
quantum state. However, one can duplicate a set of basis states of
$I$, for instance, the qubit states up and down ($|1\rangle$
and $|0\rangle$). One can also discuss an approximate, optimized
copying of a general linear combination of the basis states of
$I$ [3-5]. A added limitation of these copying procedures has been that the
{\it initial\/} state of $C$ (or more generally, of
the systems which are imprinted with the copies) must be {\it fixed}.
This feature makes it unlikely that any interesting interference
effects will be involved in the copying process.

Here we propose to explore those quantum-mechanical processes
that do not involve any restriction on the initial
state of the target system(s), 
even though the property of making copies will
be meaningful only for the basis states of the input system $I$.
If we require that the basis states of $I$ at time $t$ be copied in
such a way that both $I$ and $C$, and if needed, another copy $D$,
are all in that basis state at time $t+\Delta t$ for an
arbitrary initial state of $C$ (and $D$), then
one can easily verify that no unitary transformation can accomplish
the desired mapping. Such quantum copying is not possible.

Our proposal is to consider instead the process in which an initial
state of $I$, from the basis set $|1\rangle$, $|0\rangle$, is
duplicated in {\it at least two\/} of the three final states $I$,
$C$, $D$.
Thus, we consider three two-state systems. The initial state of $I$,
as long as it is one of the qubit states, will be ``multiplied'' in
such a way that at time $t+\Delta t$ two or three of the systems $I$,
$C$, $D$, are in that state, but we do not know if it is two or
three, and in the case of two, which two are in that state.
A unitary quantum evolution is possible that satisfies these
conditions; we provide an explicit example. We note that the same unitary
operator will also ``evolve'' an arbitrary linear combination
of the basis states of $I$. However, the resulting state does not
involve any exact copies of that linear combination. 

Quantum copying has applications in quantum cryptography and signal
transmission---a field in which presently theoretical and first
experimental results are available [7-23]. It can also find uses
in quantum computing, reviewed, e.g., in [24-33]. These fields deal with 
quantum dynamical processes that involve ``binary'' states constructed from
the up and down states of two-state systems (qubits), such as photon
polarization states or
spin-$1\over 2$ quantum states. We will use the terms
``qubit'' or ``spin.''  Study of coherent quantum evolution is also
of great ``basic science'' value.

The outline of the rest of this work is as follows. In the
rest of this section, we define our
``blind fanout'' copying model. In Section 2, an explicit Hamiltonian is derived for
the three-qubit system involved in the process. It turns out that the Hamiltonian 
involves three-spin interactions. Therefore, in Section 3, we
also derive a reduction of
the copying process in terms of a sequence of two-spin and one-spin ``gates'' in a 
formulation popular in the quantum-computing literature [24-27]. These gates must be 
applied in sequence by switching the interactions on
and off. Sections 3 also includes 
a summarizing discussion of our results.

Let us label the states of the combined
system $I+C+D$ by $|111\rangle$,
$|110\rangle$, $|101\rangle$, $|100\rangle$, $|011\rangle$,
$|010\rangle$, $|001\rangle$, $|000\rangle$, where the order of the
systems is $|ICD\rangle$. One can then argue that unitary $8\times
8$ matrices can be found that accomplish the desired transformation.
The requirement is
that any linear combination of the states
$|1CD\rangle$ is mapped onto a linear combination of
$|111\rangle$, $|110\rangle$, $|101\rangle$ and $|011\rangle$, while
any linear combination of the states $|0CD\rangle$ is mapped
onto a linear combination of
$|100\rangle$, $|010\rangle$, $|001\rangle$ and $|000\rangle$.
The general unitary transformation actually has many free parameters;
it is by no means limited or special. Many different quantum
evolutions accomplish the task.

For our explicit calculations we choose the simplest root to the
desired copying: we consider a unitary transformation that flips (and
possibly changes phases of) the basis states only in the subspace
of $|100\rangle$, $|011\rangle$. The $8\times 8$ unitary evolution
matrix $U$ can then be represented as follows:

$$ U=\pmatrix{{\cal I}_{3\times3}& & \cr
 & {\cal U}_{2\times 2}& \cr
 & & {\cal I}_{3\times3} } \;\; . \eqno(1) $$

\NI Here $\cal I$ are unit matrices. The subscripts indicate matrix
dimensions while all the undisplayed elements are zero. The most general
form of the matrix $\cal U$ is 

$$ {\cal U}=\pmatrix{0 & e^{i\beta} \cr
 e^{i\alpha}&0 } \;\; . \eqno(2) $$

\ 

\NI {\bf 2. Derivation of the Hamiltonian}

\ 

Our aim is to calculate the Hamiltonian $H$ according to

$$ U=e^{-iH\Delta t / \hbar} \;\; . \eqno(3) $$

\NI We adopt the usual approach in the quantum-computing
literature [24-32]
of assuming that the (constant)
Hamiltonian $H$ ``acts'' during the time 
interval $\Delta t$, i.e., we only consider evolution from $t$ to 
$t+\Delta t$. The dynamics can be externally timed, with $H$ 
being switched on at $t$ and off
at $t+\Delta t$. The time interval $\Delta t$ is then related to the 
strength of couplings in $H$ which are of order $\hbar/\Delta t$. One
can replace the constant Hamiltonian $H$ by $f(t)H$ provided the
shape or ``protocol'' function $f(t)$ averages to 1 over the time
interval $\Delta t$.
This allows for a smoother time dependence [33] without the need
to introduce time-ordering in (3).

To obtain an expression for $H$, we calculate the ``logarithm'' of $U$ in
its diagonal representation. One can verify that the diagonalizing matrix
$T$, such that $T^\dagger U T$ is diagonal, is of the same structure
as $U$ in (1), with the nontrivial part $\cal U$ replaced by $\cal T$, where

$$ {\cal T}={1 \over \sqrt{2}}
\pmatrix{e^{i\beta /2} & e^{i\beta /2} \cr
 e^{i\alpha /2} & - e^{i\alpha /2} }
\;\; . \eqno(4) $$

\NI In the diagonal representation, the Hamiltonian is
the diagonal $8\times 8$ matrix $-\hbar A
/ \Delta t$, where $A$ has diagonal elements
$2\pi N_1$, $2\pi N_2$, $2 \pi N_3$, ${1\over 2}(\alpha
+\beta)+2\pi N_4$, ${1\over 2}(\alpha
+\beta)+\pi + 2\pi N_5$, $2\pi N_6$, $2\pi N_7$, $2\pi N_8$.
Here $N_j$ are arbitrary integers. 

The Hamiltonian
is then obtained as $H=-\hbar T A T^\dagger / \Delta t$, and
it depends on the two (real) parameters $\alpha$ and $\beta$
and on the integers $N_j$.
We restrict the number of parameters to obtain a specific
example. In fact, we seek a Hamiltonian with few energy gaps [33].
However, we would also like to have a symmetric energy level structure.
The following choice leads to a particularly elegant result for $H$.
We put $N_j=0$ for $j=1,2,3,6,7,8$, and
also $\alpha+\beta+\pi+2\pi(N_4+N_5)=0$ and $N_5-N_4=N$.
This corresponds to the following energies: $E_{1,2,3}=0$, $E_4=
\pi\hbar \left(N+{1\over 2}\right)/\Delta t$, $E_5=-E_4$, $E_{6,7,8}=0$.

The resulting Hamiltonian 
depends only on one real parameter,

$$ \gamma = (\alpha-\beta) / 2 \;\; , \eqno(5) $$

\NI and on one arbitrary integer, $N$. All the diagonal elements of the
Hamiltonian will be zero with these choices of parameters. Indeed,
calculation of $H$ yields the result that this $8\times 8$
matrix with elements $H_{mn}$, where $m$ labels the rows and
$n$ the columns, has only two nonzero entries, 

$$ H_{45} = {\pi\hbar \over \Delta t} \left(N+{1\over 2}\right)
e^{-i\gamma } \;\;\; {\rm and} \;\;\;
H_{54} = {\pi\hbar \over \Delta t} \left(N+{1\over 2}\right)
e^{i \gamma } \;\; . \eqno(6) $$

Any matrix in a space with a multiple-qubit basis can be expanded in
terms of the direct products of the four ``basis'' \ $2\times 2$
matrices for each of the two-level systems involved: the unit matrix
$\cal I$, and the standard Pauli matrices $\sigma_x$, $\sigma_y$,
$\sigma_z$. The latter are proportional to spin components for
two-state systems which are the spin states of spin-$1\over 2$
particles. We will use the spin-component nomenclature, and their
representation in terms of the Pauli matrices. We report here the
result of such an expansion for the Hamiltonian $H$. While its matrix
form is simple and only contains two nonzero elements, the
spin-component representation is surprisingly complicated,

$$ \eqalign{ H &= {\pi \hbar \over 4 \Delta t} 
\left(N+{1\over 2}\right)\cr &\times\Big[(\cos \gamma )
\big( \sigma_{xI}\sigma_{xC}\sigma_{xD} - 
 \sigma_{xI}\sigma_{yC}\sigma_{yD} +
 \sigma_{yI}\sigma_{xC}\sigma_{yD} +
 \sigma_{yI}\sigma_{yC}\sigma_{xD} \big)\cr
&\hphantom{y}-(\sin \gamma )
\big( \sigma_{yI}\sigma_{yC}\sigma_{yD} -
 \sigma_{yI}\sigma_{xC}\sigma_{xD} +
 \sigma_{xI}\sigma_{yC}\sigma_{xD} +
 \sigma_{xI}\sigma_{xC}\sigma_{yD} \big) \Big]
\;\; .} \eqno(7) $$

\ 

\NI {\bf 3. Reduction in Terms of Quantum Gates, and Discussion}

\ 

We note that the Hamiltonian (7) involves three-spin interactions.
The triplet $x,y$-component products are 
essential in the GHZ-paradox in quantum mechanics [34,35]. However, in
that case these operators are {\it measured}. In fact, the need for
multispin interactions in the Hamiltonian
is a shortcoming as far as actual
realizations, for instance, in the field of quantum computing, are
concerned. Indeed, two-spin interactions are much more common and
better understood theoretically and experimentally in solid-state
and other systems, than three-spin interactions.

As mentioned earlier, our choice of the Hamiltonian is not
unique. Its simplicity in the matrix form has allowed exact
analytical result (7) be obtained. We have also explored certain unitary
transformation choices more general than (1). However, 
presently we cannot offer a quantum signal splitting process of the type
proposed in this work that
can be accomplished ``in one shot'' with two-spin interactions only.

There are results in the quantum-computing literature [36-39]
that establish that any unitary transformation in a multiqubit space
can in principle be represented with arbitrary high accuracy
by a {\it sequence\/} of
two-spin and one-spin ``quantum gates'' which implies at most two-spin interactions;
these interactions must be switched on and off sequentially. 
However, generally the number of such gates involved may be quite large,
and no systematic
``reduction'' procedure seems to follow from the existence-type proofs [36-39].
For our copying process, though, we managed to obtained a reduction,
basically by guessing the gate sequence. 

For simplicity, we put $\alpha=\beta=0$ so that our unitary matrix defined in (1) and
(2) only contains elements 1 or 0. A quantum-gate sequence that generates this
unitary transformation is shown in Figure~1. In involves the standard
quantum-computing NOT and controlled-controlled-NOT (CCNOT) gates [24-27]. The CCNOT
gate is also know as Toffoli 
gate. It corresponds to the binary function whereby the NOT is applied on the
``controlled'' qubit (denoted by $\oplus$ in the figure) only when
both ``controlling'' qubits
(denoted by $\bullet$) are 1. Its quantum-computing version is still
a three-spin gate. However,
it can be expressed in terms of the two-spin controlled-NOT (CNOT)
and single-spin-rotation
quantum gates, e.g., [39]. We point out that explicit Hamiltonians for
single-spin rotations and for CNOT are, respectively, one-spin and two-spin,
and they have been considered in the literature, e.g., [6,33,40]. 
  
In summary, we proposed a variant of the quantum
copying/signal splitting
in which the initial state is multiplied but there is
uncertainty in which of the two-state systems involved is the
multiple copy stored. In our scheme the initial
copy-system states are not fixed. 

The authors wish to thank R. Cleve for help with identifying the utility of
CCNOT gates in guessing the decomposition shown in Figure~1. They also wish to
acknowledge instructive discussions with C.H. Bennett, A. Ekert, M.
Hillery, S.P. Hotaling, and L.S. Schulman.
This work has been supported in part by US Air
Force grants, contract numbers F30602-96-1-0276 and F30602-97-2-0089. 
This financial assistance is gratefully acknowledged.

\NP

\centerline{\bf FIGURE CAPTION}

\ 

\noindent\hang {\bf Figure~1}:\ \ Reduction of the unitary transformation (1)-(2),
with $\alpha=\beta=0$, to a sequence of NOT and CCNOT gates; see text for details.

\NP
 
\centerline{\bf REFERENCES}{\frenchspacing
 
\item{[1]} W.K. Wooters and W.H. Zurek, Nature {\bf 299}, 802 (1982). 

\item{[2]} D. Dieks, Phys. Lett. {\bf 92} A, 271 (1982). 

\item{[3]} V. Bu\v zek and M. Hillery, Phys. Rev. A {\bf 54}, 1844 (1996).

\item{[4]} M. Hillery and V. Bu\v zek, ``Quantum Copying:
Fundamental Inequalities'' (preprint quant-ph/9701034). 

\item{[5]} V. Bu\v zek, S.L. Braunstein, M. Hillery and D. Bruss,
``Quantum Copying: A Network'' (preprint quant-ph/9703046). 

\item{[6]} D. Mozyrsky, V. Privman and M. Hillery, Phys.
Lett. A {\bf 226}, 253 (1997).

\item{[7]} A. Ekert, Nature {\bf 358}, 14 (1992).

\item{[8]} C.H. Bennett, G. Brassard and A. Ekert, Scientific
American, October 1992, p. 26.

\item{[9]} C.H. Bennett, F. Bessette, G. Brassard, L. Savail and J.
Smolin, J. Cryptology {\bf 5}, 3 (1992).

\item{[10]} C.H. Bennett, G. Brassard, C. Cr\' epeau, R. Jozsa, A.
Peres and W.K. Wooters, Phys. Rev. Lett. {\bf 70}, 1895 (1993).

\item{[11]} A. Muller, J. Breguet and N. Gisin, Europhys. Lett. {\bf
23}, 383 (1993).

\item{[12]} P.D. Townsend, J.G. Rarity and P.R. Tapster, Electron.
Lett. {\bf 29}, 1291 (1993).

\item{[13]} P.D. Townsend, Electron. Lett. {\bf 30}, 809 (1993).

\item{[14]} A.K. Ekert, B. Huttner, M.G. Palma and A. Peres,
Phys. Rev. A {\bf 50}, 1047 (1994).

\item{[15]} B. Huttner and A. Peres, J. Mod. Opt. {\bf 41}, 2397 (1994).

\item{[16]} B. Huttner and A.K. Ekert, J. Mod. Opt. {\bf 41}, 2455 (1995).

\item{[17]} J.D. Franson and B.C. Jacobs, Electron. Lett. {\bf 31}, 232 (1995).

\item{[18]} A. Muller, H. Zbinden and N. Gisin, Europhys. Lett. {\bf 33}, 335 (1996).

\item{[19]} J.I. Cirac, P. Zoller, H.J. Kimble and H. Mabuchi,
Phys. Rev. Lett. {\bf 78}, 3221 (1997).

\item{[20]} B. Huttner, N. Imoto, N. Gisin and T. Mor, Phys. Rev. A
{\bf 51}, 1863 (1995).

\item{[21]} A. Peres, ``Unitary Dynamics for Quantum Codewords''
(preprint quant-ph/9609015).

\item{[22]} T. Mor, ``Reducing Quantum Errors and Improving Large Scale
Quantum Cryptography'' (preprint quant-ph/9608025).

\item{[23]} H. Zbinden, J.D. Gautier, N. Gisin, B. Huttner,
A. Muller and W. Tittel, Electron. Lett. (to appear).

\item{[24]} C.H. Bennett, Physics Today, October 1995, p. 24. 

\item{[25]} D. Deutsch, Physics World, June 1992, p. 57.

\item{[26]} D.P. DiVincenzo, Science {\bf 270}, 255 (1995). 

\item{[27]} A. Ekert and R. Jozsa, Rev. Mod. Phys. {\bf 68}, 733 (1996).

\item{[28]} S. Haroche and J.-M. Raimond, Physics Today, August 1996, p. 51.

\item{[29]} R. Landauer, {Philos. Trans. R. Soc. London Ser.} 
A {\bf 353}, 367 (1995). 

\item{[30]} S. Lloyd, Science {\bf 261}, 1563 (1993). 

\item{[31]} A. Steane, Appl. Phys. B {\bf 64} 623 (1997).

\item{[32]} B. Schwarzschild, Physics Today, March 1996, p. 21. 

\item{[33]} D. Mozyrsky, V. Privman and S.P. Hotaling, Int. J. Modern Phys. B (to
appear).

\item{[34]} D.M. Greenberger, M. Horne and A. Zeilinger, in ``Bell's
Theorem, Quantum Theory, and Conceptions of the Universe,'' M.
Kafatos, editor (Kluwer, Dordrecht, 1989), p. 69. 

\item{[35]} N.D. Mermin, Physics Today, June 1990, p. 9.

\item{[36]} A. Barenco, Proc. R. Soc. Lond. A {\bf 449}, 679 (1995).

\item{[37]} D.P. DiVincenzo, Phys. Rev. A {\bf 51}, 1015 (1995).

\item{[38]} S. Lloyd, Phys. Rev. Lett. {\bf 75}, 346 (1995).

\item{[39]} A. Barenco, C.H. Bennett, R. Cleve, D.P. DiVincenzo,
N. Margolus, P. Shor, T. Sleator, J.A. Smolin and H. Weinfurter,
Phys. Rev. A {\bf 52}, 3457 (1995).

\item{[40]} I.L. Chuang and Y. Yamamoto, ``The Persistent Quantum Bit''
(preprint quant-ph/9604030).

}

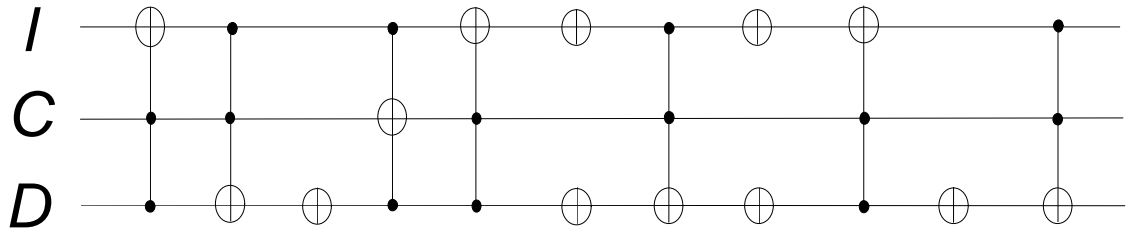
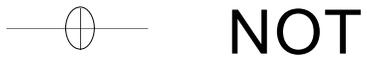
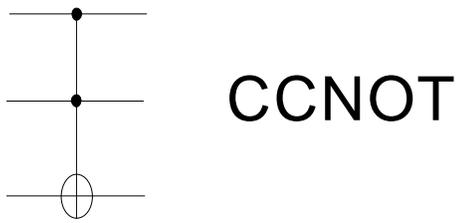
\bye